\documentclass[twocolumn,showpacs,%superscriptaddress,
preprintnumbers,amsmath,amssymb,prl]{revtex4-2}

\usepackage{graphicx}% Include figure files
\usepackage{dcolumn}% Align table columns on decimal point
\usepackage{bm}% bold math
\usepackage{float}
\usepackage{CJK}
\usepackage{ulem}
\usepackage[dvipsnames]{xcolor}  % 加载 dvipsnames 才能用 NavyBlue
\definecolor{DeepBlue}{rgb}{0.0, 0.0, 0.6}  % 可调为更深或更浅
\usepackage[colorlinks,linkcolor=DeepBlue,urlcolor=DeepBlue,citecolor=DeepBlue]{hyperref}

\newcommand{\bs}{\boldsymbol}

\renewcommand{\vec}{\boldsymbol}

\usepackage{tikz,xcolor,hyperref}

\definecolor{lime}{HTML}{A6CE39}
\DeclareRobustCommand{\orcidicon}{
	\begin{tikzpicture}
	\draw[lime, fill=lime] (0,0) 
	circle [radius=0.16] 
	node[white] {{\fontfamily{qag}\selectfont \tiny ID}};
	\draw[white, fill=white] (-0.0625,0.095) 
	circle [radius=0.007];
	\end{tikzpicture}
	\hspace{-2mm}
}
\foreach \x in {A, ..., Z}{%
	\expandafter\xdef\csname orcid\x\endcsname{\noexpand\href{https://orcid.org/\csname orcidauthor\x\endcsname}{\noexpand\orcidicon}}
}
\foreach \x in {A, ..., Z}{%
	\expandafter\xdef\csname orcid\x\endcsname{\noexpand\href{https://orcid.org/\csname orcidauthor\x\endcsname}{\noexpand\orcidicon}}
}

\begin{document}

\title{Dual-polarization structure and nuclear structure effect on $\Lambda$ polarization}% in $^{16}$O + $^{197}$Au at $\sqrt{s_{\rm NN}} = 7.7$ GeV}
%Dual Rotational Polarization of $\Lambda$-Hyperons Induced by Nuclear Clustering}
%(Or {\color{blue}{Dual polarization `swirls' and nuclear structure effect on $\Lambda$ polarization}})

\begin{CJK*}{UTF8}{gbsn}

\author{Xian-Gai Deng(邓先概)\orcidA{}}
\email{xiangai\_deng@fudan.edu.cn}
\affiliation{Key Laboratory of Nuclear Physics and Ion-beam Application (MOE), Institute of Modern Physics, Fudan University, Shanghai 200433, China}
\affiliation{Shanghai Research Center for Theoretical Nuclear Physics， NSFC and Fudan University, Shanghai 200438, China}

\author{Yu-Gang Ma(马余刚)\orcidB{}}
\email{mayugang@fudan.edu.cn}
\affiliation{Key Laboratory of Nuclear Physics and Ion-beam Application (MOE), Institute of Modern Physics, Fudan University, Shanghai 200433, China}
\affiliation{Shanghai Research Center for Theoretical Nuclear Physics， NSFC and Fudan University, Shanghai 200438, China}

\date{\today}

\begin{abstract}
%Within the Ultra-relativistic Quantum Molecular Dynamics framework, we investigate the spin polarization of $\Lambda$ hyperons in $^{16}$O + $^{197}$Au collisions at $\sqrt{s_{\rm NN}} = 7.7$ GeV, with a focus on nuclear structural effects arising from different $\alpha$-cluster configurations in $^{16}$O, including chain, kite, square, and tetrahedral arrangements, as well as a spherical reference. For the first time, a  clear dual rotational polarization signal is observed  in central $^{16}$O + $^{197}$Au collisions. The polarization further exhibits a pronounced structure-dependent behavior across centralities, particularly in the backward rapidity region. In peripheral collisions, the local polarizations $P_x$ and $P_y$ for the square and tetrahedral configurations display angular distributions opposite to those found in the spherical case. These findings shed light on the interplay between vorticity and spin in relativistic collisions, suggesting a promising experimental approach to exploring nuclear clustering at various facilities.

We report a novel manifestation of spin-vorticity interplay in relativistic heavy-ion collisions. Using $^{16}$O+$^{197}$Au at $\sqrt{s_{\rm NN}}=7.7$ GeV as a test case, we show that the $\Lambda$ hyperon exhibits a clear dual-polarization structure, observed here in central $^{16}$O + $^{197}$Au collisions for the first time. The polarization is further  highly sensitive to the intrinsic nuclear geometry: different $\alpha$-cluster configurations of $^{16}$O, ranging from chain-like to tetrahedral, lead to distinct polarization patterns across centralities. In particular, the backward rapidity region and peripheral events display striking structure-dependent variations, including opposite angular distributions of local polarizations $P_x$ and $P_y$ compared with a spherical reference. These findings reveal that nuclear clustering leaves measurable imprints on hyperon spin alignment in relativistic collisions. Our results open a promising avenue for probing nuclear structure in short-lived systems and highlight a new spin-sensitive mechanism relevant for upcoming experiments at RHIC and future facilities.

\end{abstract}


\maketitle
\end{CJK*}

%\label{introduction}

{\it Introduction---} Spin is a fundamental property of particles in quantum mechanics. In relativistic heavy-ion collisions, Liang and Wang proposed in 2005 that parton scatterings can polarize quarks along the same direction via spin-orbital coupling, leading to observable effects such as hyperon polarization~\cite{LZT_2005} and vector meson spin alignment~\cite{Liang2}. These predictions were later confirmed by $\Lambda$ polarization measurements at RHIC-STAR, which revealed a global polarization decreasing with increasing collision energy and indicated extremely strong vorticity in the quark-gluon plasma, up to $10^{21}\mathrm{s}^{-1}$~\cite{STAR1_2017,STAR:2018gyt}. More recently, global spin alignment of $\phi$ mesons was reported~\cite{STAR_2023spin}, further confirming the global polarization of quark matter.

These discoveries have sparked intense interest in spin phenomena in heavy-ion collisions (see, for example, Refs.~\cite{KL_2017,LH17,Shi:2017wpk,Beca1,Xia:2018tes,Wu:2019eyi,WDX19,XYL_2020,YBI_2020,LiuYC-2020,FBecattini-2020,BFu-2021-0,Chen1,Chen2,XGHuang-2025,Sun1,Liu,Chen3,Liu1}). There are still many open questions regarding the energy dependence of hyperon polarization~\cite{LH17}, the observed $\Lambda$-$\bar{\Lambda}$ polarization splitting ~\cite{LH17,STAR-Centrality-2024}, behavior near the production threshold ~\cite{XGDeng2020,XGDeng2022,Yu-B-Ivanov2022} and the so-called ``spin sign puzzle'' associated with local polarization ~\cite{FBecattini2018,BCFu-2021}. While hydrodynamic and transport models generally reproduce global polarization~\cite{Shen:2020mgh,Wu:2021xgu,Karpenko:2021wdm}, discrepancies persist at low beam energies 
($\sqrt{s_{\rm NN}} < 7.7$ GeV), HADES and STAR observe a continued rise in polarization, reaching 5\% at 3 GeV ~\cite{HADES-2022,STAR-2021-SNN3GeV}, which contrasts with some model predictions of a turnover ~\cite{XGDeng2020,Yu-B-Ivanov2021}. Furthermore, recent isobar collisions (Ru+Ru and Zr+Zr) reveal no significant difference in the polarization of the $\Lambda$ and $\bar{\Lambda}$ particles, with the latest results suggesting that the difference is further diminishing ~\cite{STAR-Centrality-2024}. Local polarization components, such as $P_z$, exhibit unexpected azimuthal patterns ~\cite{STAR-Pzpolarization-2019}, indicating missing elements in current theoretical descriptions. Shear-induced effects have been proposed to explain these features~\cite{BCFu-2021}, and toroidal vortex structures have also been explored in light-heavy nucleus collisions~\cite{MALisa2021}.

%It is well established that some $\alpha$-conjugate light nuclei exhibit cluster structures~\cite{KIkeda-1968,WOerzten-2001,VVovchenko-2017,ILombardo-2023}, with Giant Dipole Resonance (GDR) studies providing evidence for $\alpha$-cluster configurations in $^{12}$C and $^{16}$O~\cite{WBHe-2014}. Most nuclear clustering research focuses on low- to intermediate-energy regimes~\cite{KIkeda-1968,WOerzten-2001,VVovchenko-2017,ILombardo-2023}. In 2014, Broniowski and Arriola proposed that relativistic heavy-ion collisions could serve as a novel probe of these cluster structures~\cite{WBroniowski-2014}. Subsequent work has increasingly highlighted the role of initial geometric fluctuations in such collisions~\cite{SZhang-2017,Li2020,YYWang-2024,Gia}. Recently, the STAR Collaboration's precise measurements of nuclear shape in $^{238}$U+$^{238}$U collisions have further demonstrated the potential of relativistic collisions for probing nuclear structure~\cite{CJZhang-2024,Giu,Jia}.

Separately,  $\alpha$-conjugate light nuclei are known to exhibit cluster structures ~\cite{KIkeda-1968,WOerzten-2001,WBHe-2014,VVovchenko-2017,ILombardo-2023,Ye1,Ye2}.
%with giant dipole resonance studies providing evidence for $\alpha$-clusters in $^{12}$C and $^{16}$O~\cite{WBHe-2014}.
While most clustering research focuses on the low- to intermediate-energy regimes, relativistic heavy-ion collisions have recently emerged as a novel means of probing nuclear structure (see, for example, Refs.~\cite{WBroniowski-2014,SZhang-2017,Li2020,YYWang-2024,Li,Jia2025}). Precise measurements in collisions involving $^{238}$U have demonstrated the sensitivity of relativistic collisions to nuclear shape ~\cite{CJZhang-2024,Giu,Jia}.

Building on these insights, a key question remains: can distinct nuclear configurations generate characteristic vorticity patterns and spin polarization in relativistic collisions? To address this issue, we are investigating the effects of vorticity and polarization linked to cluster structures in light-heavy nucleus collisions at relativistic energies. This research will provide new insights into nuclear clustering properties and their manifestation in spin phenomena.

{\it Model and method for the calculation---}In this work, different $\alpha$-cluster configurations of $^{16}$O are introduced as projectile nuclei into the Ultra-relativistic Quantum Molecular Dynamics (UrQMD) model, a well-established transport framework widely used to simulate relativistic heavy-ion collisions~\cite{SA98,MB99,HP08,JS18}. It incorporates essential physics mechanisms such as hadronic rescattering, color string fragmentation, and the production and decay of resonances. Further details can be found in Refs.~\cite{HP08}. UrQMD has demonstrated strong predictive power across a wide range of collision energies, from BNL AGS energies ($E_{\text{lab}} = 1$--10 A GeV), through CERN SPS energies (20--160 A GeV), up to top RHIC energies ($\sqrt{s_{\text{NN}}} = 200$ GeV), and even into the LHC domain (up to 2.76 TeV for Pb+Pb collisions)~\cite{MM09,PPB10,SS19}.

The spin polarization can be given by the mean spin vector. For a fermion with mass $m$ and spin $s$=$\frac{1}{2}$, $\frac{3}{2}$, $\cdots$, it can be written as~\cite{Becattini:2013fla,Fang:2016vpj,Liu:2020flb,XGDeng2022,VVorronyuk2025},
\begin{equation}
S^{\mu} = -\frac{s(s+1)}{6m}(1-n_F)\epsilon^{\mu\nu\rho\sigma}p_{\nu}\varpi_{\rho\sigma},             
\label{SPINTENSOR-0}
\end{equation}
where $n_F = n_F(x,p)$ and $\varpi_{\mu\nu}=\frac{1}{2}(\partial_{\nu}\beta_{\mu}-\partial_{\mu}\beta_{\nu})$ are the Fermi-Dirac function and thermal vorticity tensor, respectively. One can write the thermal vorticity tensor as, $\bs{\varpi}_{T}=\frac{1}{2}\Big{[} \nabla \Big{(}\frac{\gamma}{T}\Big{)}+\partial_{t}\Big{(}\frac{\gamma \bs{v}}{T}\Big{)}\Big{]}$, $\bs{\varpi}_{S}=\frac{1}{2} \nabla \times \Big{(}\frac{\gamma \bs{v}}{T} \Big{)}$,
where $\bs{\varpi}_{T} $ and $\bs{\varpi}_{S}$ are called `$T$' and `$S$' thermal vorticity, respectively. In thermal vorticity tensor, $\beta^\mu = \beta u^\mu$ with $\beta = 1/T$ the inverse temperature and $u^\mu = \gamma(1,\bs{v})$ is the fluid four-velocity with $\gamma = 1/\sqrt{1-\boldsymbol{v}^2}$ the Lorentz factor. And the four-velocity satisfies the normalization condition: $u^\mu u_{\nu}$ = 1. The velocity field is calculated as described in Ref.~\cite{WTDeng2016,XGDeng2020,XGDeng2022}, $\bs{v}({\bf x}) = \sum_{i}^{N} ({\bf p}_i/{E}_i)\rho({\bf x},{\bf x}_i)/\sum_{i}^{N} \rho({\bf x},{\bf x}_i)$, where $\vec{p}_i$ and $E_i$ are the momentum and energy of the $i$-th particle located at $\bf{x}_i$, $N$ is the total particle number, and $\rho({\bf x},{\bf x}_i)$ is a smearing function. As in Ref.~\cite{HP08}, the smearing function can be written as,
\begin{equation}
\begin{split}
 &\rho({\bf x},{\bf x}_i) =  \\
 &\frac{\gamma_{z}}{(2\pi\sigma^2)^{3/2}}\,{\rm exp} \Big{\{} -\frac{({\rm x}-{\rm x}_{i})^{2}+({\rm y}-{\rm y}_{i})^{2}+[\gamma_{z}({\rm z}-{\rm z}_{i})]^{2}}{2\sigma^{2}}\Big{\}},                  
\end{split}
\label{edensity}
\end{equation}
where $\sigma = 1.48$ fm for baryons~\cite{Hartnack:1997ez} and $\sigma = 0.98$ fm for mesons from a constituent quark number scaling for volume $\sigma_{ \pi} = (2/3)^{1/3}\sigma_{\rm p,n}$ as
done in Refs.~\cite{XGDeng2020,XGDeng2022}. Furthermore, temperature $T$ serves as a measure of the energy density $\epsilon$, via the relation $\epsilon = a T^{4}$. The prefactor $a$ is chosen here to be $a = \pi^{2}(16+10.5N_{f})/30 $ with $N_{f} = 3$. By solving the eigen equation, $T^{\mu}_{\nu}u^{\nu} = \epsilon u^{\mu}$, one can obtain the energy density. The $T^{\mu\nu}$ is energy-momentum tensor, which can be defined as
\begin{equation}
T^{\mu\nu}({\bf x}) = \sum_{i}\frac{p_{i}^{\mu}p_{i}^{\nu}}{E_{i}}\rho({\bf x},{\bf x}_i), %\nonumber 
\label{SPINTENSOR-Tmunu-2}
\end{equation}
where $p_{i}^{\mu} = (E_{i}, \vec{p}_{i})$ denotes the four-momentum of the $i$-th particle. For the polarization of $\Lambda$ or $\bar\Lambda$ hyperon, typically one transforms Eq.~(\ref{SPINTENSOR-0}) from c.m. frame of nucleus-nucleus collision to the rest frame of $\Lambda$ or $\bar\Lambda$ by a Lorentz boost (Here is a vector form):
\begin{equation}
\bs{S}^{\ast}({x},{\bf p}) = \bs{S}-\frac{{\bf p} \cdot \bs{S}}{E_{p}(m+E_{p})}{\bf p}, 
\label{SPINTENSOR-3}
\end{equation}
where $E_{p}$, $p$ and $m$ are the $\Lambda$'s energy, momentum, mass in the center of mass (c.m.) frame of nucleus-nucleus collision. The averaged spin vector is given by (using energy density $\epsilon({{\bf x}})$ as a weight), $\langle \bs{S}^{\ast} \rangle = \big{[}\sum_{j}^{N} \bs{S}^{\ast}({\bf x}_{j},{\bf p}) \epsilon({\bf x}_{j})\big{]}/\big{[}\sum_{j}^{N} \epsilon({\bf x}_{j}) \big{]}$,  where the summation is over all $\Lambda$ or $\bar{\Lambda}$ within a given kinematic region and ${\bf x}_j$ is the freeze-out coordinate of the $j$-th $\Lambda$ or $\bar{\Lambda}$. Applying this algorithm to ${\Lambda}$ or $\bar{\Lambda}$ and averaging over all played-out collision events, one can get the averaged global polarization in the three-direction $\bs{n}$,
\begin{equation}
P_{\bs{n}} = 2 \langle \bs{S}^{\ast} \rangle \cdot \bs{n}.                    
\label{SPINTENSOR-4}
\end{equation}
It is worth noting that a global equilibrium of the spin degree of freedom is assumed in relating polarization to the thermal vorticity. The $\Lambda$ polarization is calculated at the freeze-out moment.

\begin{figure}[htb]
%\begin{figure}
\setlength{\abovecaptionskip}{0pt}
\setlength{\belowcaptionskip}{8pt}
\includegraphics[scale=1.05]{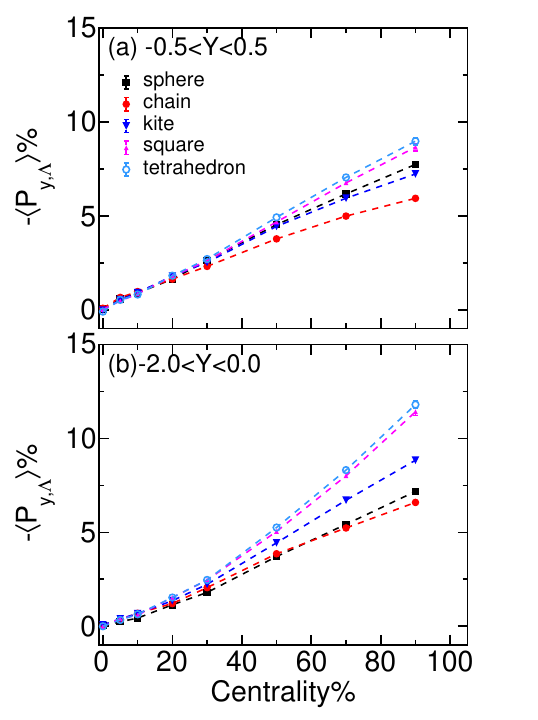}
\caption{Global $\Lambda$ polarization (${P}_{y}$) as a function of centrality in $^{16}$O + $^{197}$Au collisions at $\sqrt{s_{\mathrm{NN}}} = 7.7$ GeV for 0.2 $< {p_T}<$ 3.0 GeV/$c$, shown in (a) the midrapidity region $-0.5 < {Y} < 0.5$ and (b) the backward rapidity region $-2.0 < {Y} < 0.0$.}
\label{fig:fig1}
\end{figure}

%\begin{figure}[htb]
%%\begin{figure}
%\setlength{\abovecaptionskip}{0pt}
%\setlength{\belowcaptionskip}{8pt}
%\includegraphics[scale=1.0]{./figures/Fig3-centrality-4.pdf}
%\caption{(Color online) The same as Fig.~\ref{fig:fig1} but for the difference between $\Lambda$ and $\bar{\Lambda}$.}
%\label{fig:fig3}
%\end{figure}

%\begin{figure}[htb]
%\begin{figure}
%\setlength{\abovecaptionskip}{0pt}
%\setlength{\belowcaptionskip}{8pt}
%\includegraphics[scale=1.02]{./figures/Fig4-PT-3-0.3.pdf}
%\caption{(Color online) $\Lambda$ polarization ($\rm{P}_{y}$) as a function of $\rm{p_T}$ in $^{16}$O + $^{197}$Au collisions at $\sqrt{s_{\mathrm{NN}}} = 7.7$ GeV, shown in (a) the full rapidity region and (b) the midrapidity region $-0.5 < \rm{Y} < 0.5$ at centrality = 30\%.}
%\label{fig:fig4}
%\end{figure}

\begin{figure*}[htb]
%\begin{figure}
\setlength{\abovecaptionskip}{0pt}
\setlength{\belowcaptionskip}{8pt}
\includegraphics[scale=0.72]{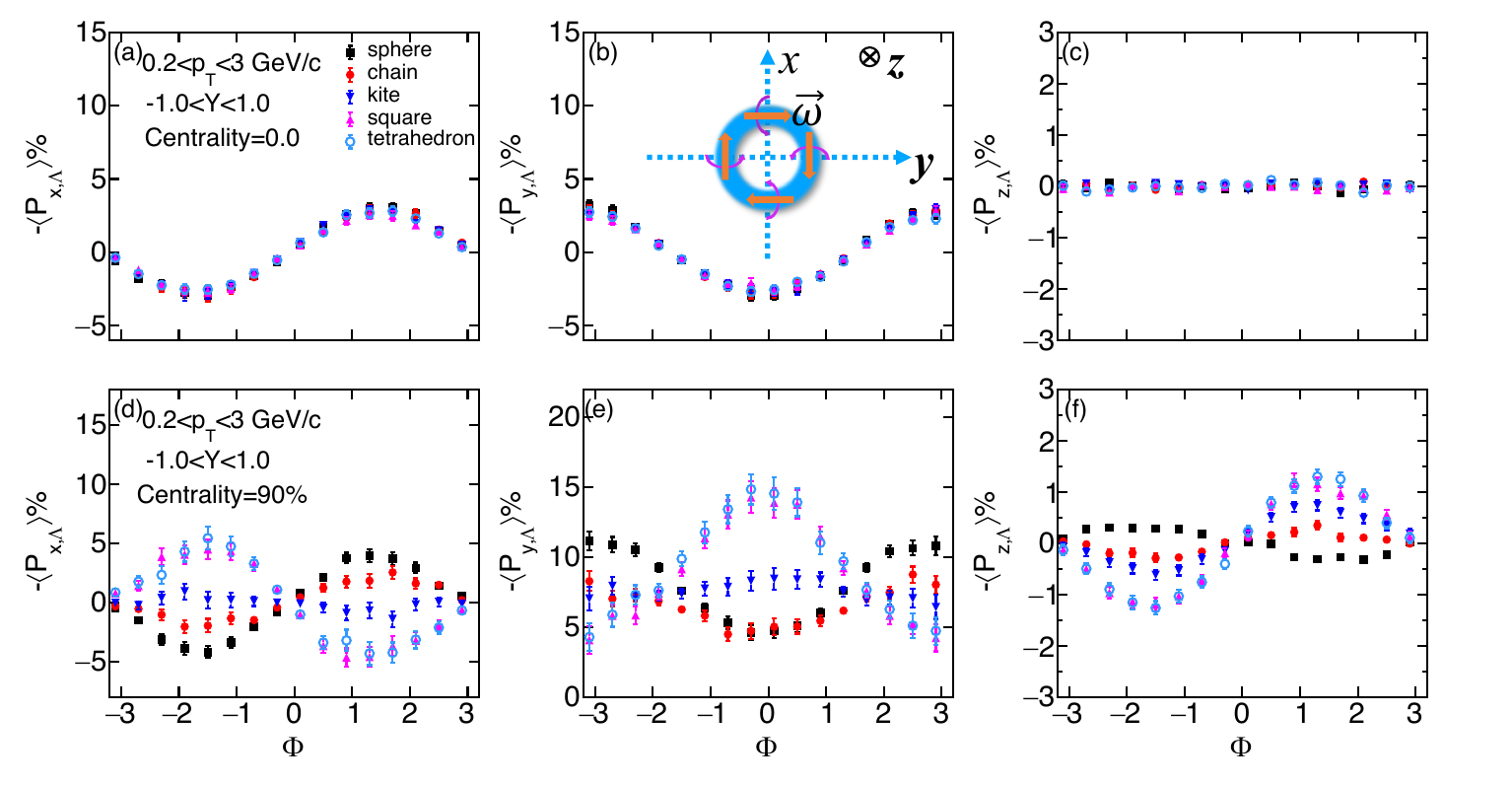}
\caption{The polarization components ${P}{x}$, ${P}{y}$, and ${P}{z}$ as functions of azimuthal angle $\Phi$ in $^{16}$O + $^{197}$Au collisions at $\sqrt{s_{\mathrm{NN}}} = 7.7$ GeV, with $0.2 < p_T < 3.0$ GeV/$c$ and $-1.0< Y < 1.0$. Panels (a)--(c) show the results for centrality $=0\%$, and (d)--(f) show the results for  centrality $=90\%$.}
\label{fig:fig2}
\end{figure*}

{\it Results and discussion---}
%\label{resultsSH}
Figure~\ref{fig:fig1} shows the global $\Lambda$ polarization as a function of centrality, calculated as described in the previous section. $P_y$ is averaged over all $\Lambda$ hyperons applying a rapidity ($Y=\frac{1}{2}[(E+p_{z})/(E-p_{z})]$) selection of $|Y| < 0.5$ or $-2.0 < Y < 0.0$, together with a transverse momentum cut of $0.2 < p_T < 3.0$ GeV/$c$ in $^{16}$O + $^{196}$Au collisions. Several $\alpha$-cluster configurations of the projectile $^{16}$O nucleus are considered, including spherical, chain, kite, square, and tetrahedral arrangements~\cite{WBHe-2014}. Centrality is defined as $C = b^2/b_{\rm max}^2$, with $b_{max} = R_{T} + R_{P} = 1.15A_{T}^{1/3} + 1.15A_{P}^{1/3}$.
As shown in Fig.\ref{fig:fig1}(a), $P_y$ rises with centrality, reaching up to $\sim 90\%$, larger than that in $^{197}$Au + $^{197}$Au collisions~\cite{XGDeng2022,GY_2019,STAR-Centrality-2024}, consistent with previous observations that smaller systems can produce higher polarization~\cite{GY_2019,SAlzhrani2022}. Differences among configurations are small at low centrality but become pronounced at high centrality. In the target-like rapidity region $-2.0 < {Y} < 0.0$, a clear hierarchy emerges: tetrahedron $>$ square $>$ kite $>$ sphere $>$ chain. Feed-down contributions are not included in this analysis.

\begin{figure}[htb]
%\begin{figure}
\setlength{\abovecaptionskip}{0pt}
\setlength{\belowcaptionskip}{8pt}
\includegraphics[scale=0.88]{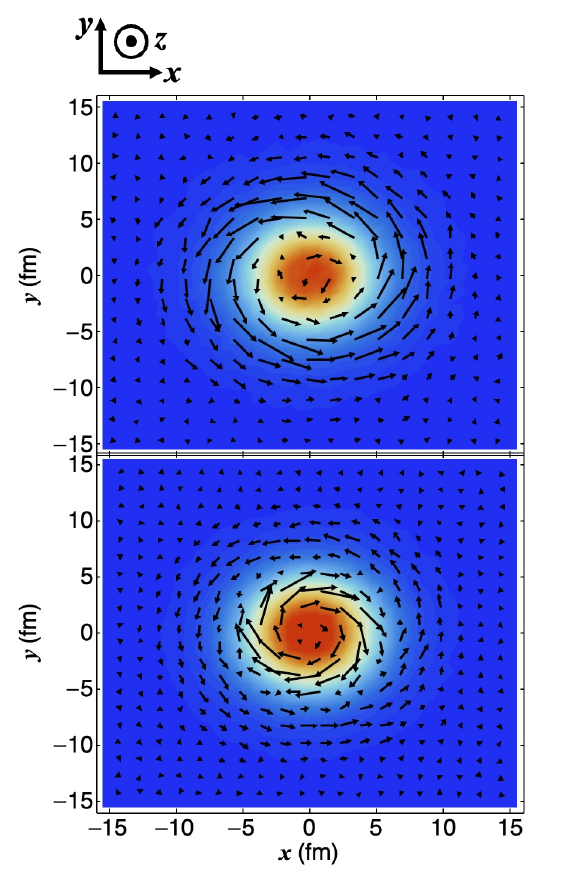}
\caption{The transverse polarization fields ($P_{x}$ and $P_{y}$) of the $\Lambda$  in the $x$-$y$ plane for $^{16}$O+$^{197}$Au collisions at $\sqrt{s_{\rm{NN}}} = 7.7$ GeV. The results are shown with a transverse momentum cut of $0.2 <p_{T} < 3.0$ GeV/$c$ and rapidity cut of $-1 <{Y} < 1$(upper), and without kinematic cuts (lower).}
\label{fig:fig3}
\end{figure}

\begin{figure*}[htb]
%\begin{figure}
\setlength{\abovecaptionskip}{0pt}
\setlength{\belowcaptionskip}{8pt}
\includegraphics[scale=0.94]{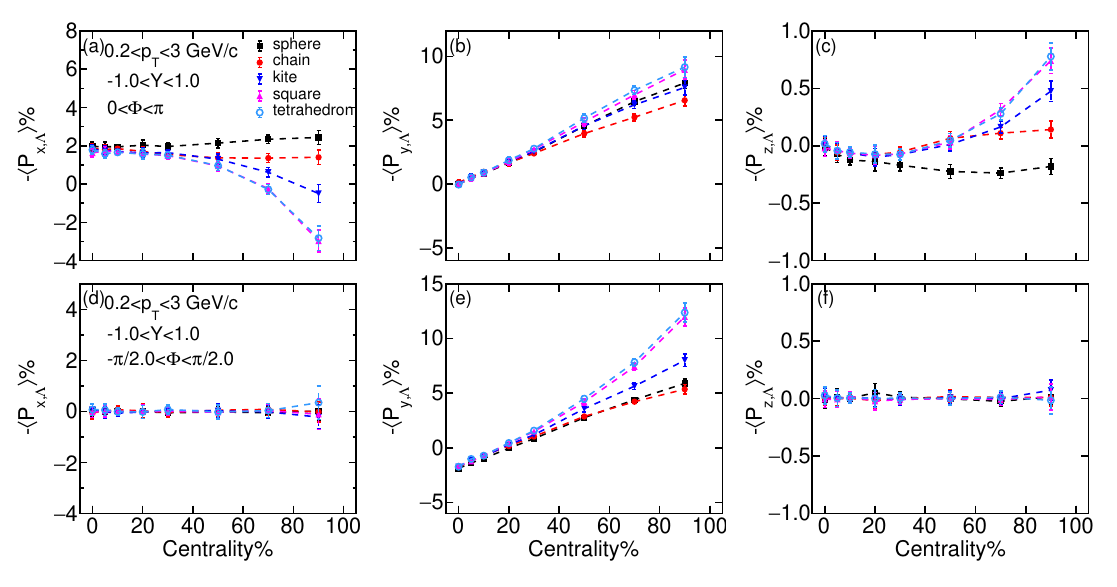}
\caption{The polarization components ${P}{x}$, ${P}{y}$, and ${P}{z}$ as functions of centrality in $^{16}$O + $^{197}$Au collisions at $\sqrt{s_{\mathrm{NN}}} = 7.7$ GeV, for $0.2 < p_T < 3.0$ GeV/$c$ and $-1 < Y < 1$. Panels (a)--(c) correspond to $0 < \Phi < \pi$, while (d)--(f) correspond to $-\pi/2 < \Phi < \pi/2$.}
\label{fig:fig4}
\end{figure*}

The azimuthal distribution of $\Lambda$ polarization, referred to as local polarization, is shown in Fig.~\ref{fig:fig2}. In central collisions (0$\%$ centrality), $-\langle P_{x,\Lambda} \rangle$ exhibits a sinusoidal pattern over $2\pi$, while $-\langle P_{y,\Lambda} \rangle$ follows a cosine-like dependence, indicating the formation of a vortex-ring structure~\cite{MALisa2021}. As illustrated in the inset of Fig.~\ref{fig:fig2} (b), particle flow (purple arrows) induced by $^{16}$O traversing $^{197}$Au generates vorticity (orange arrows). Along the positive $x$ axis, vorticity leads to positive polarization along {\rm y} (with a negative sign convention), whereas in the negative $x$ direction ($\Phi \sim \pi$ or $-\pi$) polarization is negative. Similarly, $-\langle P_{x,\Lambda} \rangle$ reaches minima and maxima at $\Phi \sim -\pi/2$ and $\pi/2$, respectively (Fig.~\ref{fig:fig2} (a)). To visualize this more intuitively, Fig.~\ref{fig:fig3} presents the distribution of transverse polarization of $\Lambda$ in the $x$--$y$ plane. The red-blue contours denote the freeze-out positions of $\Lambda$, while the arrows indicate the polarization directions and relative magnitudes. With $0.2 < p_{T} < 3.0$ GeV/$c$ and $-1 <{Y} < 1$ (upper panel), a polarization swirl emerges around the participant periphery, consistent with a vortex-ring structure, whereas polarization at the center vanishes. Interestingly, an oppositely oriented polarization swirl appears in the projectile-penetration region when no momentum or rapidity cuts are applied (lower panel). When we decompose the thermal vorticity into $S$ and $T$ components (as mention above), we find that the $S$ component of the vorticity leads to counterclockwise polarization, while the $T$ component leads to clockwise polarization. This indicates that the observed dual-polarization structure originates from the distinct contributions of the $S$ and $T$ components. In the peripheral region, the vorticity is dominated by the $S$ component ($\bs{\varpi}_{S}$), corresponding to the spatial vorticity term; this is caused by the outward flow of target matter as the projectile penetrates, forming a vortex ring that produces peripheral counterclockwise polarization. In the central region, the vorticity is dominated by the $T$ component ($\bs{\varpi}_{T}$), corresponding to the acceleration term; the clockwise polarization arises from expansion. We note that the spin vector can be written as $\bs{S}=\frac{1}{4m}\left(E_{p}\bs{\varpi}_{S}+{\bf p}\times \bs{\varpi}_{T}\right)$. From this expression, it is clear that the acceleration term is determined by the cross product of $\mathbf{p}$ and $\bs{\varpi}_T$. This also explains why, in the lower panel of Fig.~\ref{fig:fig3}, a clockwise polarization emerges in the central region when particles from a large rapidity interval (with large $p_{z}$) are considered. And since this polarization originates from the acceleration term, the associated vorticity cannot be strictly regarded as a vortex ring. As a cross-check, when we only consider $0.2 < p_{T} < 3.0$ GeV/$c$ and $(Y < -1 || Y > 1)$, $-\langle P_{x,\Lambda} \rangle$ and $-\langle P_{y,\Lambda} \rangle$ have opposite signs compared with Fig.~\ref{fig:fig2}(a) and (b). This also implies that within $(Y < -1 || Y > 1)$, there exists another polarization structure. 

%As shown in Fig.~\ref{fig:fig2}(c), longitudinal component $-\langle P_{z,\Lambda} \rangle$ is nearly zero, indicating weaker transverse expansion. All components are symmetric, resulting in vanishing global polarization, and no significant differences are observed among different cluster configurations in this regime. In contrast, for peripheral collisions (90\% centrality), distinct differences emerge among various cluster configurations in the distributions of all polarization components. For instance, in Fig.~\ref{fig:fig2}(d), $-\langle P_{x,\Lambda} \rangle$ for the spherical configuration remains similar to that in central collisions. However, for clustered $^{16}$O configurations such as the tetrahedron, $-\langle P_{x,\Lambda} \rangle$ exhibits an opposite phase in its angular distribution-clearly indicating structural effects. Similar behavior is also seen in $-\langle P_{y,\Lambda} \rangle$. For $-\langle P_{z,\Lambda} \rangle$, configurations such as the square and tetrahedron yield finite values with a sine-like angular dependence, suggesting that cluster-structured nuclei generate distinct vorticity patterns in the transverse plane. The influence of cluster structure manifests not only in the magnitude but also in the distribution shape of polarization. 

As shown in Fig.~\ref{fig:fig2} (c), the longitudinal component $-\langle P_{z,\Lambda} \rangle$ is nearly zero, reflecting weak transverse expansion. All components remain symmetric, leading to vanishing global polarization with no visible differences among cluster configurations. In sharp contrast, peripheral collisions (90$\%$ centrality) exhibit a pronounced configuration dependence across all polarization components. For example, in Fig.~\ref{fig:fig2} (d), $-\langle P_{x,\Lambda} \rangle$ for the spherical case resembles that in central events, whereas clustered configurations, such as the tetrahedron, show an opposite phase in the angular distribution--clear evidence of structural effects. A similar trend appears in $-\langle P_{y,\Lambda} \rangle$. Moreover, in $-\langle P_{z,\Lambda} \rangle$, square and tetrahedral configurations yield finite sine-like oscillations, indicating that clustered nuclei generate distinct vorticity patterns in the transverse plane. Thus, the cluster structure affects not only the magnitude but also the angular shape of the polarization.

%These effects can be understood as a consequence of how nuclear geometry impacts the initial conditions of the collision. Cluster configurations introduce localized density enhancements and geometric asymmetries, which seed nonuniform velocity gradients and vorticity patterns in the initial fireball. For example, a tetrahedral arrangement leads to anisotropic energy deposition in the transverse plane, enhancing shear flow along specific directions. These local velocity shears translate into spatially modulated vorticity, which through spin-orbit coupling, results in the observed azimuthal variation and sign reversals of local polarization. In contrast, spherical nuclei produce smoother initial profiles, yielding more symmetric and weaker vorticity structures.

These effects arise from the impact of nuclear geometry on the initial collision conditions. Clustered configurations create localized density hot spots and geometric asymmetries, seeding nonuniform velocity gradients and vorticity in the fireball. A tetrahedral arrangement, for instance, induces anisotropic energy deposition in the transverse plane, amplifying shear flow along preferred directions. Such local velocity shears generate spatially modulated vorticity which, via spin--orbit coupling, leads to the observed azimuthal variations and sign reversals of polarization. By contrast, spherical nuclei yield smoother initial profiles and consequently more symmetric, weaker vorticity fields.

%To further quantify the effects, Fig.~\ref{fig:fig4} shows the centrality dependence of the $\Lambda$ polarization components within the mid-rapidity range ($-1 < \rm{Y} < 1$) and transverse momentum interval ($0.2 < p_T$ [GeV/$c$] $< 3.0$), under two different azimuthal angle selections. In Fig.~\ref{fig:fig4}(a), for $0 < \Phi < \pi$, the spherical nucleus shows an almost constant polarization value with negligible centrality dependence. In contrast, the kite, square, and tetrahedron configurations exhibit a clear centrality-dependent trend and even sign changes in peripheral collisions--further supporting the structural origin. For the $-\langle P_{y,\Lambda} \rangle$ component shown in Fig.~\ref{fig:fig4}(b), all configurations display increasing magnitudes with centrality. In Fig.~\ref{fig:fig4}(c), $-\langle P_{z,\Lambda} \rangle$ remains close to zero for the spherical case, but increases significantly for clustered cases, indicating enhanced transverse-plane vorticity generation. In Figs.~\ref{fig:fig4}(d)--(f), for $-\pi/2 < \Phi < \pi/2$, symmetry ensures that both $-\langle P_{x,\Lambda} \rangle$ and $-\langle P_{z,\Lambda} \rangle$ vanish at all centralities. However, $-\langle P_{y,\Lambda} \rangle$ exhibits trends similar to those in Fig.~\ref{fig:fig4}(b), maintaining a finite negative value even in central collisions and showing larger configuration-dependent differences at higher centralities.

To quantify these effects, Fig.~\ref{fig:fig4} presents the centrality dependence of $\Lambda$ polarization components at mid-rapidity ($-1<Y<1$) and $0.2<p_{T}<3.0$ GeV/$c$, under two azimuthal angle selections. For $0<\Phi<\pi$ [Figs.~\ref{fig:fig4} (a)--(c)], the spherical case shows nearly constant values with little centrality dependence, while clustered configurations (kite, square, tetrahedron) exhibit clear trends and even sign reversals in peripheral events--highlighting structural effects. $-\langle P_{y,\Lambda} \rangle$ grows in magnitude with centrality, and $-\langle P_{z,\Lambda} \rangle$ stays near zero for the sphere but rises for clustered cases, reflecting stronger transverse-plane vorticity. For $-\pi/2<\Phi<\pi/2$ [Figs.~\ref{fig:fig4}(d)--(f)], symmetry enforces vanishing $-\langle P_{x,\Lambda} \rangle$ and $-\langle P_{z,\Lambda} \rangle$ across centralities, whereas $-\langle P_{y,\Lambda} \rangle$ remains finite even in central collisions and shows enhanced configuration dependence at large centrality.

These findings strongly indicate that local polarization, particularly in peripheral collisions, retains imprints of the projectile nucleus's initial cluster structure, making it a promising and sensitive probe for exploring cluster configurations in light nuclei via relativistic heavy-ion collisions.

%\label{summary}
{\it Conclusions---}%To summarize, we identify for the first time a double-polarization-swirl structure in central $^{16}$O+$^{197}$Au collisions, revealed by the angular distributions of local $\Lambda$ polarization and transverse polarization fields ($P_{x}$, $P_{y}$) in the $x$--$y$ plane, providing new insight into the polarization patterns and vorticity structures in light-heavy nucleus collisions. By including different $\alpha$-cluster configurations of $^{16}$O (chain, kite, square, tetrahedron, and sphere), we find that local $\Lambda$ polarization in peripheral collisions is strongly shaped by nuclear geometry, even showing sign reversals. This constitutes the first systematic study of cluster effects on both global and local hyperon polarization. The global polarization rises with centrality, with structure-induced differences more pronounced at backward rapidity. Dependence on azimuth further reveals distinct vorticity patterns, especially in peripheral events where local polarization is highly sensitive to nuclear configuration. These results demonstrate that hyperon polarization--particularly its local components--offers a sensitive probe of nuclear clustering, and can be tested at facilities such as HADES@GSI~\cite{HADES_2009}, BM@N and NICA@JINR~\cite{VD17,MK19}, FAIR~\cite{TA17}, and HIAF~\cite{LL17}, opening new avenues to explore the spin structure of nuclear matter.
%In summary, we have identified a double-polarization swirl structure for the first time in central collisions of $^{16}$O and $^{197}$Au, as revealed by the angular distributions of local and transverse polarization fields ($P_x$, $P_y$) in the $x-y$ plane. This provides new insight into the polarization patterns and vorticity structures in light-heavy nucleus collisions. Including different $\alpha$-cluster configurations of $^{16}$O  (chain, kite, square, tetrahedron and sphere) reveals that local lambda polarization in peripheral collisions is strongly influenced by nuclear geometry, with sign reversals observed. This is the first systematic study of the effects of clusters on both global and local hyperon polarization. Global polarization increases with centrality, with structure-induced differences being more pronounced at backward rapidity. Dependence on azimuth further reveals distinct vorticity patterns, particularly in peripheral events, where local polarization is highly sensitive to nuclear configuration. These results demonstrate that hyperon polarization, particularly its local components, offers a sensitive probe of nuclear clustering and can be tested at facilities such as HADES@GSI~\cite{HADES_2009}, BM@N and NICA@JINR~\cite{VD17,MK19}, FAIR~\cite{TA17}, and HIAF~\cite{LL17}. This opens up new avenues for exploring the spin structure of nuclear matter.
In summary, we report the first identification of dual polarization swirls in central $^{16}$O+$^{197}$Au collisions, revealed by the angular distributions of the local polarization fields $P_x$ and $P_y$ in the transverse plane. This novel structure provides new insight into the connection between hyperon spin alignment and vorticity in light-heavy nuclear collisions. By comparing different $\alpha$-cluster configurations of $^{16}$O, we further demonstrate that local $\Lambda$ polarization in peripheral events is strongly shaped by nuclear geometry, including clear sign reversals relative to the spherical case. This constitutes the first systematic study of cluster effects on both global and local hyperon polarization. We find that global polarization rises with centrality, while structure-induced differences are most pronounced at backward rapidity and show distinctive azimuthal dependencies. These results establish hyperon polarization-especially its local components-as a sensitive probe of nuclear clustering, offering experimentally testable signatures at facilities such as HADES~\cite{HADES_2009}, NICA~\cite{VD17,MK19}, FAIR~\cite{TA17}, and HIAF~\cite{LL17}, and opening new directions for exploring the spin structure of nuclear matter.

{\it Acknowledgments:}
Authors thank Dr. Chen Zhong for maintaining the high-quality performance of Fudan HIRG supercomputing platform for nuclear physics. This work received partial support from the National Natural Science Foundation of China under Contract Nos. 12205049, 12147101, 12347149, 11890714, and 11925502, the Strategic Priority Research Program of CAS under Grant No. XDB34000000, the Guangdong Major Project of Basic and Applied Basic Research No. 2020B0301030008, the STCSM under Grant No. 23590780100, the Natural Science Foundation of Shanghai under Grant No. 23JC1400200, and the National Key R$\&$D Program of China No. 2023YFA1606701.

%\end{acknowledgments}


\begin{thebibliography}{99}
%Deng X.G.

\bibitem{LZT_2005} Z. T. Liang and X. N. Wang,
             \href{https://link.aps.org/doi/10.1103/PhysRevLett.94.102301}{Phys. Rev. Lett. {\bf 94}, 102301 (2005)}; Erratum, %: Globally Polarized Quark-Gluon Plasma in Noncentral Au + Au Collisions [Phys. Rev. Lett. 94, 102301 (2005)], 
             \href{https://doi.org/10.1103/PhysRevLett.96.039901}{Phys. Rev. Lett. {\bf 96}, 039901  (2006).}

\bibitem{Liang2}
            % Spin alignment of vector mesons in non-central A+A collisions
Z. T. Liang and X. N. Wang,  \href{https://doi.org/10.1016/j.physletb.2005.09.060}{Phys. Lett. B {\bf 629}, 20  (2005).}



%\bibitem{STAR1_2005} STAR Collaboration, 
 %            \href{https://doi.org/10.1016/j.nuclphysa.2005.03.085}{Nucl. Phys. A {\bf 757}, 102-183 (2005)}.
             
  \bibitem{STAR1_2017} STAR Collaboration,
             \href{https://doi.org/10.1038/nature23004}{Nature {\bf 548}, 62--65 (2017)}.
             
\bibitem{STAR:2018gyt} STAR Collaboration, 
             \href{https://doi.org/10.1103/PhysRevC.98.014910}{Phys. Rev. C \textbf{98}, 014910 (2018)}.
             
\bibitem{STAR_2023spin} STAR Collaboration,
\href{https://doi.org/10.1038/s41586-022-05557-5}{Nature {\bf 614}, 244-248 (2023)}.
%Pattern of global spin alignment of ϕ and K*0 mesons in heavy-ion collisions.             
             

\bibitem{KL_2017} Lu. Karpenko and F. Becattini, 
             \href{https://doi.org/10.1140/epjc/s10052-017-4765-1}{Eur. Phys. J. C {\bf 77}, 213 (2017)}.                                        
\bibitem{LH17} H. Li, L. G. Pang, Q. Wang {\it et al.}, 
             \href{https://doi.org/10.1103/PhysRevC.96.054908}{Phys. Rev. C {\bf 96}, 054908 (2017)}.  

\bibitem{Shi:2017wpk} S.~Shi, K.~Li and J.~Liao,
             \href{https://doi.org/10.1016/j.physletb.2018.09.066}{Phys. Lett. B \textbf{788}, 409 (2019)}.

\bibitem{Xia:2018tes} X.~L.~Xia, H.~Li, Z.~B.~Tang  {\it et al.},
            \href{https://doi.org/10.1103/PhysRevC.98.024905}{Phys. Rev. C \textbf{98}, 024905 (2018)}.
            
\bibitem{Wu:2019eyi} H.~Z.~Wu, L.~G.~Pang, X.~G.~Huang {\it et al.},
            \href{https://doi.org/10.1103/PhysRevResearch.1.033058}{Phys. Rev. Research. \textbf{1}, 033058 (2019)}. 

\bibitem{WDX19}D. X. Wei, W. T. Deng, X. G. Huang, 
            \href{https://link.aps.org/doi/10.1103/PhysRevC.99.014905}{Phys. Rev. C {\bf 99}, 014905 (2019)}.       


\bibitem{Beca1}F. Becattini, I. Karpenko, M. A. Lisa, {\it et al.},
\href{https://doi.org/10.1103/PhysRevC.95.054902}{Phys. Rev. C {\bf 95}, 054902 (2017)}.
%Global hyperon polarization at local thermodynamic equilibrium with vorticity, magnetic field, and feed-down, 

%\bibitem{Beca2}F. Becattini, J. Liao, M. A. Lisa,
%       \href{https://doi.org/10.1007/978-3-030-71427-7}{Lecture Notes in Physics, Vol. {\bf 987}, Springer, 2021.}
 %Strongly Interacting Matter Under Rotation. 
 
\bibitem{XYL_2020} Y. L. Xie, D. J. Wang, L. P. Csernai,
            \href{https://doi.org/10.1140/epjc/s10052-019-7576-8}{Eur. Phys. J. C {\bf 80}, 39 (2020)}.  
           
\bibitem{YBI_2020} Yu. B. Ivanov and A. A. Soldatov,
            \href{https://doi.org/10.1103/PhysRevC.102.024916}{Phys. Rev. C {\bf 102}, 024916 (2020)}.


\bibitem{LiuYC-2020} Y. C. Liu and X.~G.~Huang,
        \href{https://doi.org/10.1007/s41365-020-00764-z}{Nucl. Sci. Tech. \textbf{31}, 56 (2020)}.
%Anomalous chiral transports and spin polarization in heavy-ion collisions.


\bibitem{FBecattini-2020} F. Becattini and M. A. Lisa,         
            \href{https://doi.org/10.1146/annurev-nucl-021920-095245}{Annu. Rev. Nucl. Part. Sci. {\bf 70}, 395-423  (2020)}.   
%Polarization and Vorticity in the Quark–Gluon Plasma
             
\bibitem{BFu-2021-0} B.~Fu, K.~Xu, X.~G.~Huang {\it et al.},
             \href{https://doi.org/10.1103/PhysRevC.103.024903}{Phys. Rev. C \textbf{103}, 024903 (2021)}.

\bibitem{Chen1}J. Chen, Z. T. Liang, Y. G. Ma {\it et al.}, \href{https://doi.org/10.1016/j.scib.2023.04.001 }{Sci. Bull. {\bf68}, 874--877 (2023) }.
% Global spin alignment of vector mesons and strong force fields in heavy-ion collisions.


\bibitem{Chen2}J. Chen, X. Dong, Y. G. Ma {\it et al.}, \href{https://doi.org/10.1016/j.scib.2023.11.045}{ Sci. Bull. {\bf 68}, 3252--3260 (2023) }.
%Measurements of the lightest hypernucleus (HΛ3): progress and perspective.

             
\bibitem{XGHuang-2025} X.~G.~Huang,
        \href{https://doi.org/10.1007/s41365-025-01784-3}{Nucl. Sci. Tech. \textbf{36}, 208 (2025)}.


\bibitem{Sun1}K. J. Sun, D. N. Liu, Y. P. Zheng {\it et al.},
%Deciphering Hypertriton and Antihypertriton Spins from Their Global Polarizations in Heavy-Ion Collisions.
\href{https://doi.org/10.1103/PhysRevLett.134.022301}{Phys. Rev. Lett. {\bf 134}, 022301 (2025)}.

\bibitem{Liu}%From Hyperons to Hypernuclei: A New Route to Unravel Proton Spin Polarization
D. N. Liu,  Y. P.  Zheng,  W. H. Zhou {\it et al.}, \href{https://doi.org/10.48550/arXiv.2508.12193}{arxiv:2508.12193}.



\bibitem{Chen3}J. Chen, Z. T. Liang, Y. G. Ma {\it et al.}, 
\href{https://doi.org/10.1007/s11433-024-2495-1}{Sci. China Phys. Mech. Astron. {\bf 68}, 211001 (2025)}.
%Vector meson’s spin alignments in high energy reactions,

\bibitem{Liu1}R. J. Liu, J. Xu,  Y. G. Ma, 
%Spin polarization from nucleon-nucleon scatterings in intermediate-energy heavy-ion collisions.
\href{https://doi.org/10.1016/j.physletb.2025.139703}{Phys. Lett. B {\bf 868}, 139703 (2025)}.


\bibitem{STAR-Centrality-2024} STAR Collaboration,
\href{https://drupal.star.bnl.gov/STAR/blog/futong/Paper-proposal-Measurements-global-polarization-Lambda-hyperons-AuAu-collisions-RHIC-B-0}{Paper proposal: Measurements of global polarization of Lambda hyperons in Au+Au collisions from the RHIC Beam Energy Scan-II, 2024}.
             
 \bibitem{XGDeng2020} X. G. Deng, X. G. Huang, Y. G. Ma {\it et al.}, 
              \href{https://doi.org/10.1103/PhysRevC.101.064908}{Phys. Rev. C {\bf 101}, 064908 (2020)}.
                       
 \bibitem{XGDeng2022} X. G. Deng, X. G. Huang, and Y. G. Ma,
\href{https://doi.org/10.1016/j.physletb.2022.137560}{Phys. Lett. B {\bf835}, 137560 (2022).}
             
\bibitem{Yu-B-Ivanov2022} Yu. B. Ivanov and A. A. Soldatov,
\href{https://doi.org/10.1103/PhysRevC.105.034915}{Phys. Rev. C 105, 034915 (2022)}.


\bibitem{FBecattini2018} F. Becattini and Iu. Karpenko,
\href{https://doi.org/10.1103/PhysRevLett.120.012302}{Phys. Rev. Lett. {\bf120}, 012302 (2018)}.
%Collective Longitudinal Polarization in Relativistic Heavy-Ion Collisions at Very High Energy


\bibitem{BCFu-2021}B. C. Fu, S. Liu, L. G. Pang {\it et al.}, 
\href{https://doi.org/10.1103/PhysRevLett.127.142301}{Phys. Rev. Lett. {\bf127} 142301(2021)}.
%Shear-Induced Spin Polarization in Heavy-Ion Collisions
              
             

\bibitem{Shen:2020mgh} C.~Shen and L.~Yan,
             \href{https://doi.org/10.1007/s41365-020-00829-z}{Nucl. Sci. Tech. \textbf{31}, 122 (2020)}. 

\bibitem{Wu:2021xgu} S.~Wu, C.~Shen and H.~Song,
             \href{http://cpl.iphy.ac.cn/10.1088/0256-307X/38/8/081201}{Chin. Phys. Lett. \textbf{38}, 081201 (2021)}.              
\bibitem{Karpenko:2021wdm} I.~Karpenko,
           \href{https://doi.org/10.1007/978-3-030-71427-7_8}{Lect. Note Phys. {\bf 987}, 247-280 (2021)}.           
%Vorticity and Polarization in Heavy-Ion Collisions: Hydrodynamic Models          
             
 \bibitem{HADES-2022} HADES Collaboration,
           \href{https://doi.org/10.1016/j.physletb.2022.137506}{Phys. Lett. B {\bf835}, 137506 (2022)}.


\bibitem{STAR-2021-SNN3GeV}STAR Collaboration,
\href{https://doi.org/10.1103/PhysRevC.104.L061901}{Phys. Rev. C {\bf104}, L061901 (2021)}.
 %Global -hyperon polarization in Au+Au collisions at √sN N = 3 GeV,       
            
\bibitem{Yu-B-Ivanov2021} Yu. B. Ivanov,
             \href{https://doi.org/10.1103/PhysRevC.103.L031903}{Phys. Rev. C {\bf 103}, L031903 (2021)}.

      
 \bibitem{XGGou-2022} X. G. Gou for the STAR Collaboration,
        \href{https://doi.org/10.1051/epjconf/202327604007}{EPJ Web of Conferences {\bf276}, 04007 (2023)}.
                     
  
                                    
    
                 
%\bibitem{GBunce1976}G. Bunce, R. Handler, R. March {\it et al.},
%\href{https://doi.org/10.1103/PhysRevLett.36.1113}{Phys. Rev. Lett. {\bf36}, 1113 (1976)}.
             
%\bibitem{Belle-2019}Belle Collaboration,
%\href{https://doi.org/10.1103/PhysRevLett.122.042001}{Phys. Rev. Lett. {\bf122}, 042001 (2019)}.
%Observation of Transverse Λ=Λ ̄ Hyperon Polarization in e + e − Annihilation at Belle

%\bibitem{CYi-2025}C. Yi, X. Y. Wu, J. Zhu {\it et al.},
%\href{https://doi.org/10.1103/PhysRevC.111.044901}{Phys. Rev. C {\bf111}, 044901 (2025)}.
%Spin polarization of hyperons along the beam direction in p+Pb collisions at sNN = 8.16 TeV using hydrodynamic approaches



 \bibitem{STAR-Pzpolarization-2019}STAR Collaboration,
\href{https://doi.org/10.1103/PhysRevLett.123.132301}{Phys. Rev. Lett. {\bf123}, 132301 (2019)}.             
 %Polarization of Λ (Λ) Hyperons along the Beam Direction in Au+Au Collisions at √����⁢��=200  GeV

%\bibitem{Y-B-Ivanov2017}Yu. B. Ivanov and A. A. Soldatov,
%\href{https://doi.org/10.1103/PhysRevC.95.054915}{Phys. Rev. C. {\bf95} 054915 (2017)}.
 %Vorticity in heavy-ion collisions at the JINR Nuclotron-based Ion Collider fAcility
 
\bibitem{MALisa2021}M. A. Lisa, J. G. P. Barbon, D. D. Chinellato {\it et al.}
\href{https://doi.org/10.1103/PhysRevC.104.L011901}{Phys. Rev. C {\bf104}, L011901 (2021)}.
%Vortex rings from high energy central ��+�� collisions
 
\bibitem{KIkeda-1968}K. Ikeda, N. Tagikawa, and H. Horiuchi, 
      \href{https://doi.org/10.1143/PTPS.E68.464}{Prog. Theor. Phys. Supplement E. {\bf68}, 464 (1968)}.
 %The Systematic Structure-Change into the Molecule-like Structures in the Self-Conjugate 4n Nuclei 


\bibitem{WOerzten-2001}W. von Oerzten, 
      \href{https://doi.org/10.1007/s100500170052}{Eur. Phys. J. A {\bf11}, 403 (2001)}.
 %Covalently bound molecular structures in the α + 16O system

\bibitem{WBHe-2014} W. B. He, Y. G. Ma, X. G. Cao {\it et al.},
        \href{https://doi.org/10.1103/PhysRevLett.113.032506}{Phys. Rev. Lett. 113, 032506 (2014)}.
 %Giant Dipole Resonance as a Fingerprint of �� Clustering Configurations in  12 C   and  16 O  
 
\bibitem{VVovchenko-2017}V. Vovchenko, M. I. Gorenstein, L. M. Satarov {\it et al.}, 
     \href{https://doi.org/10.1007/978-3-319-44165-8}{New Horizons in Fundamental Physics, Berlin: Springer (2017)}.
     
\bibitem{ILombardo-2023}I. Lombardo, D. Dell'Aquila, 
      \href{https://doi.org/10.1007/s40766-023-00047-4}{Riv. Nuovo Cim. {\bf46}, 521--618 (2023)}.
%Clusters in light nuclei: history and recent developments

\bibitem{Ye1}K. Wei, Y. L. Ye, Z. H. Yang, %Clustering in nuclei: progress and perspectives. 
\href{https://doi.org/10.1007/s41365-024-01588-x}{Nuc. Sc. Tech. {\bf 35}, 216 (2024).} 

 \bibitem{Ye2}
 Y. L. Ye, X. F. Yang, H. Sakurai, B. S. Hu, \href{https://doi.org/10.1038/s42254-024-00782-5}{Nat. Rev. Phys. {\bf 7}, 21--37 (2025)}.

\bibitem{WBroniowski-2014} W. Broniowski and E. R. Arriola,
\href{https://doi.org/10.1103/PhysRevLett.112.112501}{Phys. Rev. Lett. {\bf112}, 112501 (2014)}.

\bibitem{SZhang-2017} S. Zhang, Y. G. Ma, J. H. Chen {\it et al.}, 
        \href{https://link.aps.org/doi/10.1103/PhysRevC.95.064904}{Phys. Rev. C {\bf  95}, 064904 (2017)}.
 %Nuclear cluster structure effect on elliptic and triangular flows in heavy-ion collisions, 

%\bibitem{SZhang-2018}S. Zhang, Y. G. Ma, J. H. Chen {\it et al.}, 
 %       \href{https://doi.org/10.1140/epja/i2018-12597-y}{Eur. Phys. J. A {\bf54}: 161(2018).}
 
\bibitem{Li2020}
%Signatures of α-clustering in 16O by using a multiphase transport model, 
Y. A.  Li, S. Zhang, Y. G.  Ma, \href{https://doi.org/10.1103/PhysRevC.102.054907 }{Phys. Rev. C {\bf 102}, 054907 (2020)}. 

% Impact of Nuclear Deformation on Relativistic Heavy-Ion Collisions: Assessing Consistency in Nuclear Physics across Energy Scales
%\bibitem{Gia}G.  Giacalone,  J. Y. Jia, C. J.  Zhang, \href{https://doi.org/10.1103/PhysRevLett.127.242301}{Phys. Rev. Lett. {\bf 127}, 242301 (2021)}.


\bibitem{YYWang-2024} Y. Y. Wang, S. Zhao, B. Cao {\it et al.}, 
         \href{https://doi.org/10.1103/PhysRevC.109.L051904}{Phys. Rev. C {\bf109}, L051904 (2024).}

\bibitem{Li}
%Benchmarking Nuclear Matrix Elements of 0⁢��⁢��⁢�� Decay with High-Energy Nuclear Collisions
Y. Li, X. Zhang, G. Giacalone, J. Yao, \href{https://doi.org/10.1103/zymp-tyjj}{Phys. Rev. Lett. {\bf 135}, 022301 (2025)}.

\bibitem{Jia2025}J. Y. Jia,
\href{https://doi.org/10.1007/s41365-025-01800-6}{Nucl. Sci. Tech. {\bf 36}, 207 (2025)}.
%Bridging nuclear physics across energy scales: from neutrinoless double‑beta decay to high‑energy heavy‑ion collisions



\bibitem{CJZhang-2024}STAR Collaboration, 
\href{https://doi.org/10.1038/s41586-024-08097-2}{Nature {\bf635}, 67--72 (2024)}.
%Imaging shapes of atomic nuclei in high-energy nuclear collisions

\bibitem{Giu} G. Giacalone, \href{https://doi.org/10.1007/s41365-024-01582-3}{Nucl. Sci.  Tech. {\bf 35}, 218 (2024)}.
%Beyond axial symmetry: high‑energy collisions unveilthe ground‑state shape of 238U

\bibitem{Jia}J. Jia, G. Giacalone, B. Bally {\it et al.}, 
%Imaging the initial condition of heavy-ion collisions and nuclear structure across the nuclide chart. 
\href{https://doi.org/10.1007/s41365-024-01589-w}{Nucl. Sci. Tech. {\bf 35}, 220 (2024). }


\bibitem{SA98} S. A. Bass, M. Belkacem, M. Bleicher {\it et al.}, 
            \href{https://doi.org/10.1016/S0146-6410(98)00058-1}{Prog. Part. Nucl. Phys. {\bf 41}, 225 (1998)}.   
\bibitem{MB99} M. Bleicher,  E. Zabrodin, C. Spieles {\it et al.}, 
            \href{https://doi.org/10.1088/0954-3899/25/9/308}{J. Phys. G {\bf 25},  1859  (1999)}.       
\bibitem{HP08} H. Petersen, J. Steinheimer, G. Burau {\it et al.},  
             \href{https://doi.org/10.1103/PhysRevC.78.044901}{Phys. Rev. C {\bf 78}, 044901 (2008)}.
                                            
\bibitem{JS18} J. Steinheimer, V. Vovchenko,  J. Aichelin {\it et al.}, 
            \href{https://doi.org/10.1051/epjconf/201817105003}{EPJ Web of Conferences {\bf 171}, 05003 (2018)}.

%\bibitem{SR80} S. R. De Groot, W. A. Van Leeuwen, and C. G. Van Weert,
%Relativistic Kinetic theory:\ Principles and Applications (North-Holland, Amsterdam, 1980).

\bibitem{MM09} M. Mitrovski, T. Schuster, G. Gr{\"a}f  {\it et al.}, 
                      \href{https://doi.org/10.1103/PhysRevC.79.044901}{Phys. Rev. C {\bf 79}, 044901 (2009)}. 

\bibitem{PPB10} P. P. Bhaduri and S. Chattopadhyay,  
                       \href{https://doi.org/10.1103/PhysRevC.81.034906}{Phys. Rev. C {\bf 81}, 034906 (2010)}. 
                       
\bibitem{SS19} S. Sombun, K. Tomuang, A. Limphirat {\it et al.}, 
                      \href{https://doi.org/10.1103/PhysRevC.99.014901}{Phys. Rev. C {\bf 99}, 014901 (2019)}.  


%\bibitem{XGDeng2025} X. G. Deng and Y. G. Ma,
 %          \href{https://doi.org/10.1103/PhysRevC.111.014904}{Phys. Rev. C {\bf 111}, 014904 (2025)}.


\bibitem{Becattini:2013fla} F.~Becattini, V.~Chandra, L.~Del Zanna {\it et al.},
  \href{https://doi.org/10.1016/j.aop.2013.07.004}{Annals Phys. \textbf{338}, 32-49 (2013)}.

\bibitem{Fang:2016vpj} R.~H.~Fang, L.~G.~Pang, Q.~Wang {\it et al.},
  \href{https://doi.org/10.1103/PhysRevC.94.024904}{Phys. Rev. C \textbf{94}, 024904 (2016)}.

\bibitem{Liu:2020flb} Y.~C.~Liu, K.~Mameda and X.~G.~Huang,
  \href{https://doi.org/10.1088/1674-1137/ac009b}{Chin. Phys. C \textbf{44}, 094101 (2020)}.
 

\bibitem{VVorronyuk2025} V. Voronyuk, N. S. Tsegelnik, and E. E. Kolomeitsev,
\href{https://link.aps.org/doi/10.1103/PhysRevC.111.034907}{Phys. Rev. C {\bf 111}, 034907 (2025).}
% title = {Hyperon global polarization in heavy-ion collisions at energies available at the JINR Nuclotron-based Ion Collider fAcility: Feed-down effects and the role of ${\mathrm{\ensuremath{\Sigma}}}^{0}$ hyperons},


 \bibitem{WTDeng2016} W.~T.~Deng and X.~G.~Huang,
            \href{https://doi.org/10.1103/PhysRevC.93.064907}{Phys. Rev. C \textbf{93}, 064907 (2016)}.


\bibitem{Hartnack:1997ez} C.~Hartnack, R.~K.~Puri, J.~Aichelin {\it et al.},
\href{https://doi.org/10.1007/s100500050045}{Eur.\ Phys.\ J.\ A {\bf 1}, 151 (1998)}.  

 
 \bibitem{GY_2019} Y. Guo, S. Z. Shu, S. Q. Feng {\it et al.}, 
\href{https://doi.org/10.1016/j.physletb.2019.134929}{Phys. Lett. B {\bf 798}, 134929 (2019)}.   
             
\bibitem{SAlzhrani2022}S. Alzhrani, S. Ryu, and C. Shen, 
\href{https://doi.org/10.1103/PhysRevC.106.014905}{Phys. Rev. C {\bf106}, 014905 (2022)}.



%\bibitem{KXu2022} K. Xun, F. Lin, A. P. Huang and M. Huang.
 %\href{https://doi.org/10.1103/PhysRevD.106.L071502}{Phys. Rev. D {\bf106}, L071502 (2022)}.
 
\bibitem{HADES_2009} HADES Collaboration, 
             \href{https://doi.org/10.1140/epja/i2009-10807-5}{Eur. Phys. J. A {\bf 41}, 243 (2009)}.
          
\bibitem{VD17} V. D. Kekelidze, V. A. Matveev, I. N. Meshkov {\it et al.},
            \href{https://doi.org/10.1134/S1063779617050239}{Phys. Part. Nucl. {\bf 48}, 727 (2017)}.                    

\bibitem{MK19} M. Kapishin (BM@N Collaboration), 
            \href{https://doi.org/10.1016/j.nuclphysa.2018.07.014}{Nucl. Phys. A {\bf 982}, 967 (2019)}.
            
\bibitem{TA17} T. Ablyazimov {\it et al.} (CBM Collaboration),
            \href{https://doi.org/10.1140/epja/i2017-12248-y}{Eur. Phys. J. A {\bf 53}, 60 (2017)}.
            
\bibitem{LL17} H. W. Zhao, H. S. Xu, G. Q. Xiao {\it et al.}, 
            \href{https://doi.org/10.1360/SSPMA-2020-0248}{Sci. Sin.-Phys. Mech. \&  Astron {\bf 50}, 112006 (2020).}
%Huizhou accelerator complex facility and its future development (in Chinese).

%=======================================================================================================
%=======================================================================================================
%=======================================================================================================
%old references============================================================================================





\end{thebibliography}
\end{document}